%% file: main.tex
\documentclass{ifacconf}

\usepackage[utf8]{inputenc}
\usepackage{graphicx}
\usepackage{natbib}
\usepackage{multirow}
\usepackage{url}
\usepackage{listings}
\usepackage{lstcustom}
\usepackage{placeins}
\usepackage{soul}

\begin{document}
\begin{frontmatter}

\title{A Conceptual Framework for Establishing Trust in Real World Intelligent Systems}

\thanks[footnoteinfo]{Work was partially supported by Research Campus of Central Hessen (Flexi Funds).}

\author[1,2]{Michael Guckert}
\author[2]{Nils Gumpfer}
\author[2]{Jennifer Hannig}
\author[3]{Till Keller}
\author[4]{Neil Urquhart}

\address[1]{Department of MND - Mathematik, Naturwissenschaften und Datenverarbeitung, Technische Hochschule Mittelhessen - University of Applied Sciences, Wilhelm-Leuschner-Stra{\ss}e 13, 61169 Friedberg, Germany, phone: +49-6031-604-452, e-mail: michael.guckert@mnd.thm.de}
\address[2]{Cognitive Information Systems, KITE - Kompetenzzentrum f\"ur Informationstechnologie, Technische Hochschule Mittelhessen - University of Applied Sciences, 61169 Friedberg, Germany}
\address[3]{Department of Internal Medicine I, Cardiology, Justus-Liebig-University Gie{\ss}en, 35390 Gie{\ss}en, Germany}
\address[4]{Edinburgh Napier University, Edinburgh EH11 4DY, United Kingdom, e-mail: n.urquhart@napier.ac.uk}

\input{abstract}


\end{frontmatter}

\section{Introduction}
\input{introduction}

\section{Case Studies}
\input{casestudies}

\section{Pattern Framework}
\input{patternframework}

\section{Related Work}
\input{relatedwork}

\section{Conclusion and Future Work}
\input{conclusion}

\begin{ack}
This project is supported by Research Campus of Central Hessen (FCMH) via Flexi Funds. The used clinical data is based on a cohort that is part of the Kerckhoff Biomarker Registry (BioReg) that is financially supported by the Kerckhoff Heart Research Institute (KHFI) and the German Center for Cardiovascular Research e.V. (DZHK). The sponsors had no influence on the study design, statistical analyses or draft of the paper. We thank Andreas Rolf and the clinical team of the Campus Kerckhoff of the Justus-Liebig-University Gie{\ss}en for help with acquisition of clinical data. We further acknowledge the support of Michael Koerber, Andreas Morgen and Bernhard Seeger regarding the ChronicleDB used for querying the ECGs. We also thank Joshua Prim for the help regarding typesetting of the final document.
\end{ack}

\bibliography{references}

\end{document}

%% file: abstract.tex
\begin{abstract}
Intelligent information systems that contain emergent elements often encounter trust problems because results do not get sufficiently explained and the procedure itself can not be fully retraced. This is caused by a control flow depending either on stochastic elements or on the structure and relevance of the input data. Trust in such algorithms can be established by letting users interact with the system so that they can explore results and find patterns that can be compared with their expected solution. Reflecting features and patterns of human understanding of a domain against algorithmic results can create awareness of such patterns and may increase the trust that a user has in the solution. If expectations are not met, close inspection can be used to decide whether a solution conforms to the expectations or whether it goes beyond the expected. By either accepting or rejecting a solution, the user's set of expectations evolves and a learning process for the users is established. In this paper we present a conceptual framework that reflects and supports this process. The framework is the result of an analysis of two exemplary case studies from two different disciplines with information systems that assist experts in their complex tasks.
\end{abstract}

%% file: introduction.tex
Human expertise in many aspects is largely based on prior knowledge and familiar patterns, which have either been acquired through personal experience or been passed on from other members of a community the individual belongs to. As long as answers to prevalent questions conform to available, well-known patterns acceptance of theses answers is not an issue. Otherwise, they must become crystallisation points of new insights and initiate the creation of new patterns that find acceptance by convincing the individual expert and the communities with their superiority. However, a stagnancy that sticks to known knowledge has long been observed and been called knowledge inertia \citep{Liao2002}.  

Today, human experts solve complex problems with instruments using algorithmic techniques that rely on emergent phenomena and thus confront humans with results that are uncommon and have been constructed in uncommon ways. Such techniques, a class to which systems that we now call intelligent systems belong to, produce results of high complexity (e.g. routes in a map with assigned vehicles for transportation of goods) or use highly complex structures to compute results (e.g. classification through a trained artificial neural network). Both situations are difficult to capture for a human being. Moreover, the construction process depends on data and self-organising algorithmic mechanisms that typically include non-deterministic elements. Solutions can be large and therefore be difficult to evaluate and validate, especially if a solution is unfamiliar to the user or the solution contains unexpected elements. Such solutions often do not conform to known patterns, neither does the way the results are produced. This quite understandably ensues traces of doubt about the validity of the result. Trust in such systems must be established as intelligent systems become a commodity in a world of increasing digitisation \citep{8558724}. Upon that, humans easily fall prey to various cognitive biases. They often tend to either overestimate their own capabilities and knowledge \citep{Pallier2002} or easily accept solutions produced by a machine without sufficient critical reflection (automation bias) \citep{Cummings04automationbias}. In this context appropriate instruments become a necessity.        

In this paper, we present two exemplary case studies in which human experts, in the first case traffic experts and in the second case doctors, use intelligent systems that produce results on the edge between known and novel patterns. Both systems use formal languages to explicitly document and formulate patterns and are able to integrate new patterns. We use the case studies to derive a conceptual framework for maintaining patterns that can be generalised for other disciplines. Before we conclude and look at future research directions we relate our ideas to current work.

%% file: casestudies.tex
In this section, we discuss two case studies that exemplify how human experts may use intelligent systems for decision support. We look at transportation planners locating micro depots and scheduling couriers and doctors, namely cardiologists, who analyse ECG recordings as a diagnostic instrument.   

\input{evolutionaryAlgorithms}
\input{cardiology}

%% file: evolutionaryAlgorithms.tex
\subsection{Case 1:	Planning Urban Logistics using Evolutionary Algorithms}

In this section we examine the use of an urban logistics problem, the micro-depot routing problem \citep{urquhart2019}, as an example knowledge domain. 
Our aim in this domain is not to solve specific problem instances, but to \textit{illuminate} the solution space in order to inform a domain expert as to the options for the use of micro-depots (MDs) and associated couriers. It is necessary  for the domain experts to have trust in the algorithm if they are to accept the solutions presented and ultimately use them to inform policy making.

\subsubsection{The Micro-Depot Problem}

The delivery of packages by courier/express parcel (CEP) companies to city centres can contribute to congestion and pollution, due to the number of delivery vehicles required. Such vehicles (traditionally light vans) pollute the atmosphere when moving and the need to park in order to deliver parcels contributes congestion. One approach to reducing the impact of CEP deliveries is to employ modes such as walking couriers, cycle couriers (cargo bikes) and electric vehicles (EVs). These modes (especially the first two) contribute far less pollution and require less parking space when making a delivery. The downside is that walking and cycle couriers have far less capacity and range. In order to effectively use such alternative couriers they may be based at MDs close to the city centre. The MD may be a locker or container positioned in a location where a conventional \textit{supply vehicle}  can deliver a supply of packages without causing congestion, the final short delivery runs between the MD and the recipient can be made by walking, cycle or EV courier. 
The use of MDs represents a major change in how goods are delivered and hence a major disruption to the associated knowledge domain. The use of MDs is effectively expanding the knowledge domain. The challenge that now exists is to assist the  expert in applying their knowledge to this expanded domain.

In the problems examined here, a range of possible MD locations have been previously identified, the solver can determine which locations are to be used.

We characterise solutions to the MD problem as follows:

 \begin{itemize}
 \item Pollution: attempt to improve air-quality in the city through lowering emissions.
 \item Distance: reduce congestion by reducing the overall distance travelled.
 \item Couriers: the number of couriers required to implement the solution.
 \item Time: the time span required to make all of the deliveries
 \end{itemize}

In addition the financial cost of the solution is used as the fitness function, financial cost being the sum of the fixed costs and costs/km for the couriers used in the solution. 

\subsubsection{Solving the Micro-Depot Problem}

\begin{figure*}
     \includegraphics[width=1.0\textwidth]{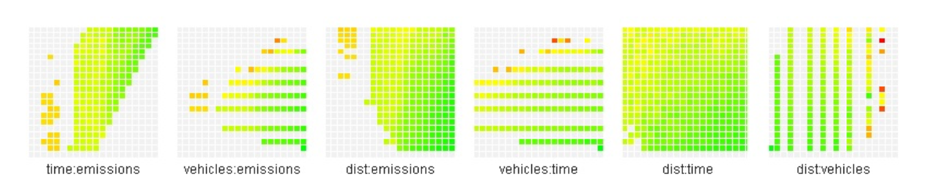}
     \caption{A visualisation of a 4-dimensional elite archive, shown as a set of 2D heat maps}
    %  , each map shows the archive from the perspective of 2 dimensions, each pair of dimensions being covered. Each individual square within each map represents a ``bin'', if it is gray no solution has been found, if it is coloured then a solution has been found, the colour is based on the fitness of the solution occupying the bin.}
     \label{fig:elite1}
 \end{figure*}

The Multi-dimensional Archive of Phenotypic Elites (MAP-Elites) is an {\it illumination algorithm} that was first introduced by \citet{Mouret-2015}. MAP-Elites creates a structured archive of high-performing solutions mapped onto solution characteristics defined by the user. A set of characteristics is identified which may be used to classify a solution (e.g. for a routing problem one might use cost, distance, delivery time span and vehicles required). Solutions are generated using mutation and recombination operators, but each solution can be classified by normalising its characteristics in order to identify a ``bin'' within the solution space that the solution belongs to. For instance we might normalise our four characteristics on a scale of 0-20, thus a solution might occupy a bin such as 5:4:2:12 for example. The number of bins in a map is calculated as $s^d$ where $s$ is the number of points on the scale and $d$ is the number of dimensions. In our example the number of bins would  be 160,000 ($20^4$).  There exists the issue as to what happens when a solution is generated that belongs to a bin that is already occupied. In this case MAP-Elites uses a {\it fitness} value to determine which solution should be allowed to occupy the bin. In our vehicle routing problem (VRP) example we could utilise distance as the fitness value and thus, when a solution is found that maps to an occupied bin, it replaces the existing solution if it represents a decrease in distance. Figure \ref{fig:elite1} shows a representation of an archive as produced by MAP-Elites based on a 4-dimensional problem as described above.

We use a representation and operators that are fully described by the authors in \citet{urquhart2019}. The representation uses a {\it grand tour} which represents the route to be taken using the supply vehicle to visit all customers, this route is constructed using the nearest-neighbour heuristic.  A chromosome is a structure which contains instructions to transfer groups of deliveries to a MD. Each gene within the chromosome represents one particular courier as follows:

\begin{center}
\begin{tabular}{|l|l|l|l|}
\hline
Tour Point & Customer Qty & MD & Courier Mode \\
\hline
\end{tabular}
\end{center}

An example gene might be :
$5,3,MD1,WALK$ which would remove the customers at positions 5,6 and 7 in the grand tour and have them delivered by a walking courier based at MD $d1$ (the 3 items removed from the grand tour are replaced by a single visit to $d1$). Mutation operators create new genes, delete old genes and randomly alter the contents of a selected gene. The recombination operator creates new chromosomes by randomly selecting genes from the parents.

In summary the chromosome is a set of instructions that may be used to convert the grand tour into a 2-tier delivery solution making using of MD and couriers. MAP-Elites provides the user with a structured set of solutions from which they may make the choice of final solution. Solutions can now be further visualised and analysed by appropriate Parallel Coordinates plots (see figure \ref{fig:pc1}).

% \begin{figure}
%     \centering
%     \includegraphics[width=0.5\textwidth]{images/Mapelites2.png}
%     \caption{A PC visualisation of a 4-dimensional elite archive.}
%     \label{fig:elite2}
% \end{figure}

% {\it  Also include a section on the "timelines" from the GECCO workshop paper. Also have a read at https://dl.acm.org/citation.cfm?id=3321897 and speak to Alexnader Hagg}

\subsubsection{Increasing Trust by Recording Algorithmic Decisions}
\label{sec:decisions}
%Ref GECCO Paper
A drawback of stochastic based meta-heuristics (including MAP-Elites) is the difficulty in explaining how a solution was arrived at. A user will have an understanding of the solution presented to them as a result of executing the algorithm, but will not have any insight as to how that solution was arrived at. 

When the solution generated does not match the user's expectation they look for an explanation as to how that solution was created.  To take an example; suppose we solve a Travelling Salesman problem using a simple constructional heuristic such as nearest neighbour (nn). The nn heuristic is simple and non-stochastic, it is possible to explain that city $A$ is followed by city $B$ as $B$ was the closest un-visited city at the point $A$ was added to the solution. The nn heuristic has the attributes of simplicity and predictability (due to being non-stochastic) which make the solutions produced easily explainable. But those attributes which make it explainable, also result in the solutions being of poor quality. It should also be remembered that VRP type problems are more complex than TSP and so simplistic heuristics such as nn are not necessarily appropriate.

A stochastic meta-heuristic (such as MAP-Elites) generates solutions by means of recombination and mutation, both of which contain random elements. Solutions are then potentially incorporated within the population using a replacement strategy \citep{holland_1992}. The random variations and chance of survival underpin the evolutionary search process. These are difficult to justify, but they are the means by which success is achieved. Explaining this process to an end-user can therefore present difficulties. \citet{hart_ross_2001} attempted to produce a visualisation of an Evolutionary Algorithm (EA) by tracking solutions and showing their antecedents. The number of operations (e.g. mutation and recombination) within an EA can make such visualisations difficult to follow and of limited value when trying to explain how the characteristics of the final solution were determined.

\begin{table*}
\tiny
\centering
\caption{The "history" of a specific bin (20:20:9:18) within the archive. Each row represents an update of the cell contents, at each point the fitness improves. The details of the change made may be ascertained by examining the solution characteristics.}
\begin{tabular}{cccccccc}
\hline
\multirow{2}{*}{\textbf{Time}} & 
\multirow{2}{*}{\textbf{Origin}} &
\multirow{2}{*}{\textbf{Updates}} & 
\multirow{2}{*}{\textbf{Fitness}} &
\multicolumn{4}{c}{\textbf{Solution Characteristics}}  
\\ \cline{5-8} 
                               &                                   &                                   &                                         & \textbf{Couriers} & \textbf{Emissions} & \textbf{Distance} & \textbf{Time} \\ \hline
5704                           & 12:4:9:11/12:12:4:17              & 0                                 & 977.11                                  & 5                 & 45.48              & 21.25             & 11.66         \\ 
5779                           & 17:16:9:19/19:20:7:9              & 1                                 & 930.21                                  & 5                 & 48.15              & 21.04             & 12.106        \\ 
14252                          & 17:16:11:18                       & 2                                 & 922.87                                  & 5                 & 42.97              & 21.19             & 10.55         \\ 
17036                          & 20:20:9:18                        & 3                                 & 922.01                                  & 5                 & 42.59              & 21.14             & 10.55         \\ 
50723                          & 11:16:9:18/12:8:7:11              & 4                                 & 919.24                                  & 5                 & 40.92              & 21.10             & 10.61         \\ 
55009                          & 11:10:9:5/19:17:7:15              & 5                                 & 881.60                                  & 5                 & 48.70              & 21.15             & 11.41         \\ 
124016                         & 13:12:9:7/18:17:9:4               & 6                                 & 880.66                                  & 5                 & 47.90              & 21.21             & 12.24         \\ 
130454                         & 15:15:11:18/14:7:7:19             & 7                                 & 878.25                                  & 5                 & 46.12              & 21.28             & 10.77         \\ 
207786                         & 19:19:11:12/15:12:11:10           & 8                                 & 878.22                                  & 5                 & 47.49              & 20.85             & 11.1          \\ 
229666                         & 10:16:7:6                         & 9                                 & 877.06                                  & 5                 & 46.58              & 20.90             & 11.06         \\ 
417443                         & 14:13:13:16/16:15:7:16            & 10                                & 875.85                                  & 5                 & 44.97              & 21.16             & 10.54         \\ 
863310                         & 18:18:4:2                         & 11                                & 875.32                                  & 5                 & 45.53              & 20.88             & 10.67         \\ 
1195610                        & 20:20:7:20                        & 12                                & 874.23                                  & 5                 & 44.88              & 20.87             & 11.00         \\ \hline
\end{tabular}
\label{tab:history}
\end{table*}

The MAP structure within MAP-Elites provides a means of identifying the "timeline" of a solution within a particular "bin". Assuming that the user has an interest in the final solution occupying a bin we can generate a history of that bin as shown in table \ref{tab:history}. It is important to emphasise that this table presents the the history of the cell and not the history of the solution. 

The table shows that 12 (update 13 represented a change to the existing solution) solutions have occupied the bin. Where two other bins are referenced then a crossover operator, followed by a mutation was used, where only 1 bin is referenced then the new solution was created by cloning followed by a mutation.  Each change results in an improvement in fitness, accompanied by some form of change in one or more characteristics. If we examine the final change we note that the number of couriers does not change but the emissions drop and the distance drops fractionally, whilst the overall time increases. We can surmise that this change is most likely a change in courier type to a lower emissions courier for a delivery run which results in less $CO_2$, but takes slightly longer.

The information used to compile table \ref{tab:history} is generated from meta-data recorded by the algorithm to a log file during executing. The meta-data records changes within the MAP structure, this provides information that can be interpreted by an algorithmic expert, e.g. by interpreting the keys contained within the origin column. Future work in this area requires the meta-data to be enhanced to include phenotype information that would allow a solution to be visualised by the domain expert showing and highlighting the differences between solutions over the timeline, the aim being to answer questions as to why particular features are or aren't present in a solution.

\subsubsection{Increasing Trust by Presenting Alternatives}
\label{sec:vis}

Expert users (e.g. logistics planners) may have greater trust in a solution, if they have been able to use their expertise within the solution construction. MAP-Elites allows  end-users to select the final solution from those presented in the archive of elites. Browsing the characteristics of such solutions and making the final selection can allow the expert user to view the range of solutions that are possible.

It can be argued that MAP-Elites is a solution space filter -- it takes the initial solution space (which is too large for the user to comprehend) and filters out a set of solutions that are representative of the solution space. In the case of the problem under discussion, the capacity of the map (assuming all bins are filled) is $20^4=160000$. From a practical perspective, it may be that MAP-Elites is  taking the massive initial solution space, and replacing it with a smaller search space that is still, from a user's perspective, too large to be of use. For instance, is asking a user to select a solution from a set of 160,000  any improvement over asking them to select a solution from the initial solution space? It becomes necessary to support the end user through the selection of  the final solution. If that support allows the end user to make a choice from thousands of alternatives then we are, perhaps, making best use of their expertise. 

\begin{figure*}
    \centering
    \includegraphics[width=0.8\textwidth]{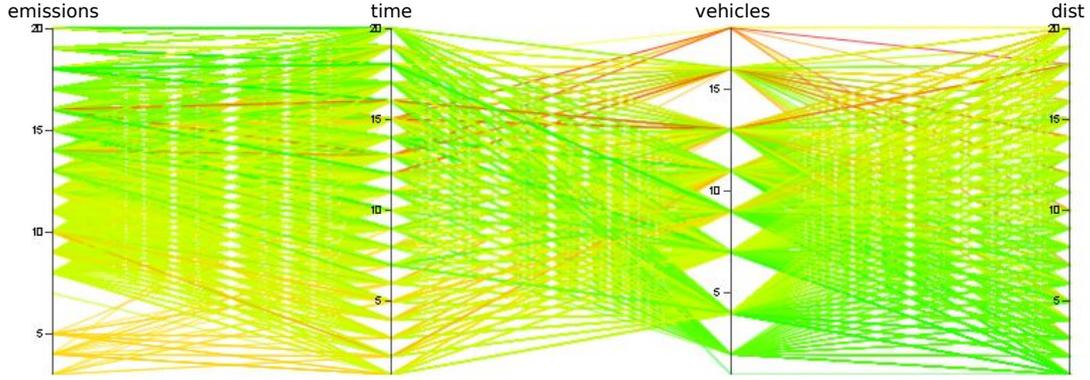}
    \caption{A Parallel Coordinate plot of a MAP of Elite solutions for a problem with 4 characteristics. }
    \label{fig:pc1}
\end{figure*}

\begin{figure*}
    \centering
    \includegraphics[width=0.8\textwidth]{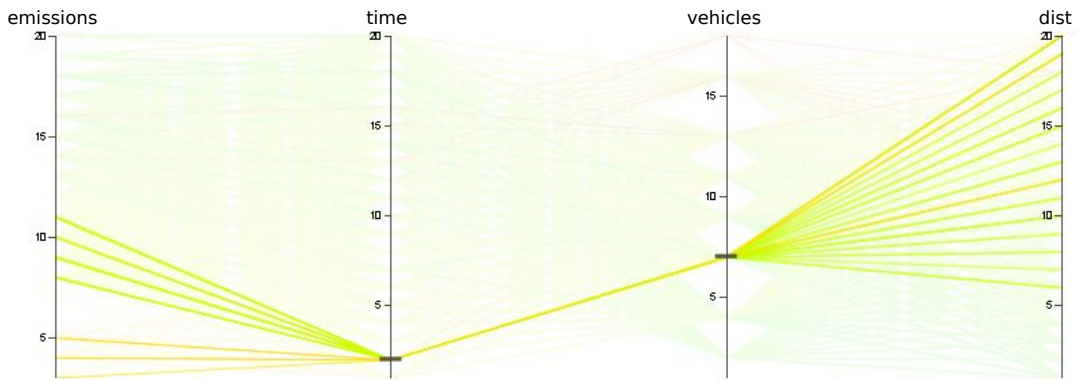}
    \caption{The plot shown in figure \ref{fig:pc1}, but with the user specifying criterion on two axes in order to highlight specific solutions.}
    \label{fig:pc2}
\end{figure*}

The end user may be supported by visualising the map of solutions using a technique such as Parallel Coordinate (PC). Within a PC plot solutions are represented as a Polyline that intersects vertical axis at the appropriate points, for a full description of the technique the reader is referred  to \citet{inselberg_shneiderman_2009}. A PC plot of the output from a typical run of MAP-Elites may be seen in figure \ref{fig:pc1}. In order for the plot to form part of a tool that supports the user it is desirable that the user interacts with the plot. The authors present the ELite VISualisation (ElVis) tool \citep{urquhart_2019el} \footnote{ElVis can be accessed at  https://commute.napier.ac.uk/upload} which allows a .CSV file containing summary details of a MAP to be visualised as a PC plot. ElVis  supports user-interaction with the PC plot. Figure \ref{fig:pc2}  demonstrates how ElVis allows the user to highlight areas of interest on the 2nd and 3rd axis (time and vehicles) leading to the selection only those solutions that match the users' criterion being highlighted. In this way the user may be guided through the the solutions found and be able to visualise alternatives as they are supported towards making their the final choice. The user has used their domain knowledge to choose a solution, rather than simply being presented with a solution by the optimisation algorithm. ElVis allows this choice to be informed, by allowing the user to gain an understanding of the solution space and what options are available to them.

\subsubsection{Using Patterns to Understand Solutions. }
\label{sec:pattern}

\begin{table*}
    \centering
    \caption{An example of a MDVRP solution  encoded in the format $<Mode>||<Depot><CustMode>$, along with an explanation of the solution.}
    %\resizebox{\textwidth}{!}{%
    \begin{tabular}{|c|l|}
    \hline
    \textbf{Encoding}   & \multicolumn{1}{c|}{VC1VC2VD1C3WC4WD3C5EC6E} \\ \hline
    \textbf{Explanation} & \begin{tabular}[c]{@{}l@{}}Customers 1 and 2 served by van from MD d1, customers 3 and 4\\  served  by walking from MD d3 etc.\end{tabular} \\ \hline
    \end{tabular}%}%
    \label{tab:exampleProb}
\end{table*}

\begin{table}
    \centering
    \caption{Examples regular user expectations, encoded as regular expressions, which may be applied to a problem encoded as per the example in table \ref{tab:exampleProb}.}
    \begin{tabular}{|l|l|}
    \hline
    \textbf{Expectation    }                             & \textbf{Regex     }                             \\ \hline
    Cust 2 is served by bike                 & C2B                                    \\ \hline
    Cust 1 is followed immediately by Cust 2 & C1\textbackslash{}DC2\textbackslash{}D \\ \hline
    Depot D2 is in use                       & ,D2\textbackslash{}D                   \\ \hline
    \end{tabular}
\end{table}

The domain knowledge and experience of an expert user will lead to them having expectations of the solution, this may  take the form of specific features that they expect to see within the solution. The confidence that the expert has in a solution may be increased if the solution contains elements that match  the expectations of the user. Assuming that the user has confidence in their own knowledge then this confidence may well extend to a solution if that solution contains features that reflect the users' knowledge. Where a solution is identified by the optimisation algorithm as being of a high quality and reflects the users' beliefs then we have match between the users' expectations and the algorithms quality measure (fitness function), we hypothesise that this will increase trust felt by the user towards the solution. When the algorithm allocates a low quality value to a solution that contains features that the user expects, then we have a mismatch between the users' model of a quality solution and that modelled within the algorithm. In this situation a number of possible outcomes exist \begin{itemize}
    \item The user can find another solution within the MAP that meets their requirements
        \item The user learns from the solution presented, accepts it and revises their expectations
    \item The fitness function is modified to closer match the users' expectations

    \item The user does not accept the presented solution and there is a breakdown in trust between the user and system
\end{itemize}

The first outcome is the most desirable and represents a specific advantage of an illumination algorithm such as MAP-Elites. The second outcome very much depends on the willingness of the user the evaluate the solutions presented by the algorithm and accept them. The third option is a longer-term answer as it may have a software development cost, but ultimately it may lead to software that produces solutions that match the expertise of the user. Machine learning may allow fitness functions to learn from the user and so adjust the weightings given to solutions that the user favours. The last option represents a failure of the optimisation process to correctly mirror the users' expectations. This may be due to incorrectly specified software or more likely a development process which was not able to solicit the end user requirements in sufficient detail.

If we take vehicle routing as an example, the user might expect to find two customers that are physically adjacent  to be adjacent within the solution. Alternatively the user might expect that a customer is serviced by particular modes, for instance a customer located within a pedestrian precinct might expect to be serviced by a courier on foot or on a bicycle.

It can be argued that the expectations of the expert user are a manifestation of the expert knowledge held by the user. Such expectations can be defined as patterns, in a similar manner to which the EA encodes potential solutions.  We can allow the user to define patterns and then note whether they exist in  solutions contained within the map. It could be argued that such patterns could be expressed as problem constraints (soft or hard as appropriate), but that would potentially bias the search and possibly leave out novel solutions that don’t include these patterns. If a pattern is not found then the user may be interested in what alternative is presented within the solution.

For the MD example we could express each solution within the map as a string of the following format:

\begin{center}
  $<Mode>||<Depot><CustMode>$
\end{center}

An encoding of the solutions using a format such as that shown above has the advantage that we can encode patterns that represent the users' expectations as regular expressions, which may be applied to the string. Providing that the users' expectations can be encoded as a regex then two opportunities present themselves:

\begin{itemize}
    \item A confidence factor may be calculated for each solution, based on the \% of regex beliefs that the solution matches.  Such a confidence factor could be used to highlight solutions within the MAP that reflect the users' beliefs. It could be argued that this value could be incorporated into a fitness function, but this needs to be treated with caution as it could lead to the evolutionary process converging on a solution that reflects the users' prejudices and views rather than one that is genuinely optimal.
    \item The regex represent the beliefs of the expert user, based on their experiences and, some cases prejudices of the user. It may be useful to highlight \textit{alternatives} that have been evolved in solutions that do not match. Suppose the user has specified that a customer should be serviced by a bicycle courier (as per the first example in table \ref{tab:exampleProb}). It might be useful to highlight solutions where the pattern is \textit{not} present. For example, if the user considers solutions that meet the overall objective of low emissions, but don't exhibit the pattern, it may become apparent that customer 2 could be served by a walking courier and still have the solution meet the users' requirements.
\end{itemize}

\subsubsection{Conclusions and Future Work}

The measures outlined above may be used to attempt to increase the confidence of expert users when dealing with real-world problems. Such users have expertise and experience that cannot necessarily be captured in an evaluation function. Making use of that expertise through the methods outlined has significant potential for not only increasing confidence but also, making use of the expert knowledge within the problem solving process. 

The timeline of solution development described in section \ref{sec:decisions} would benefit from improved interpretability by highlighting the actual differences made to the solution at each stage.  By recording the differences between the phenotype of the old and new solution. The differences can then be quantified against the changes in the fitness value and the problem characteristics. This could take the form of a visual representation (e.g. a map) of the chromosome.

The pattern matching process proposed in section \ref{sec:pattern} could be further adapted by use of a GUI to allow easier input of the patterns, rather than by specifying them as regular expressions, which may not be easily understood by the domain expert.

%% file: cardiology.tex
\subsection{Case 2: ECG Analysis}

\begin{figure*}
    \centering
    \includegraphics[width=0.75\textwidth]{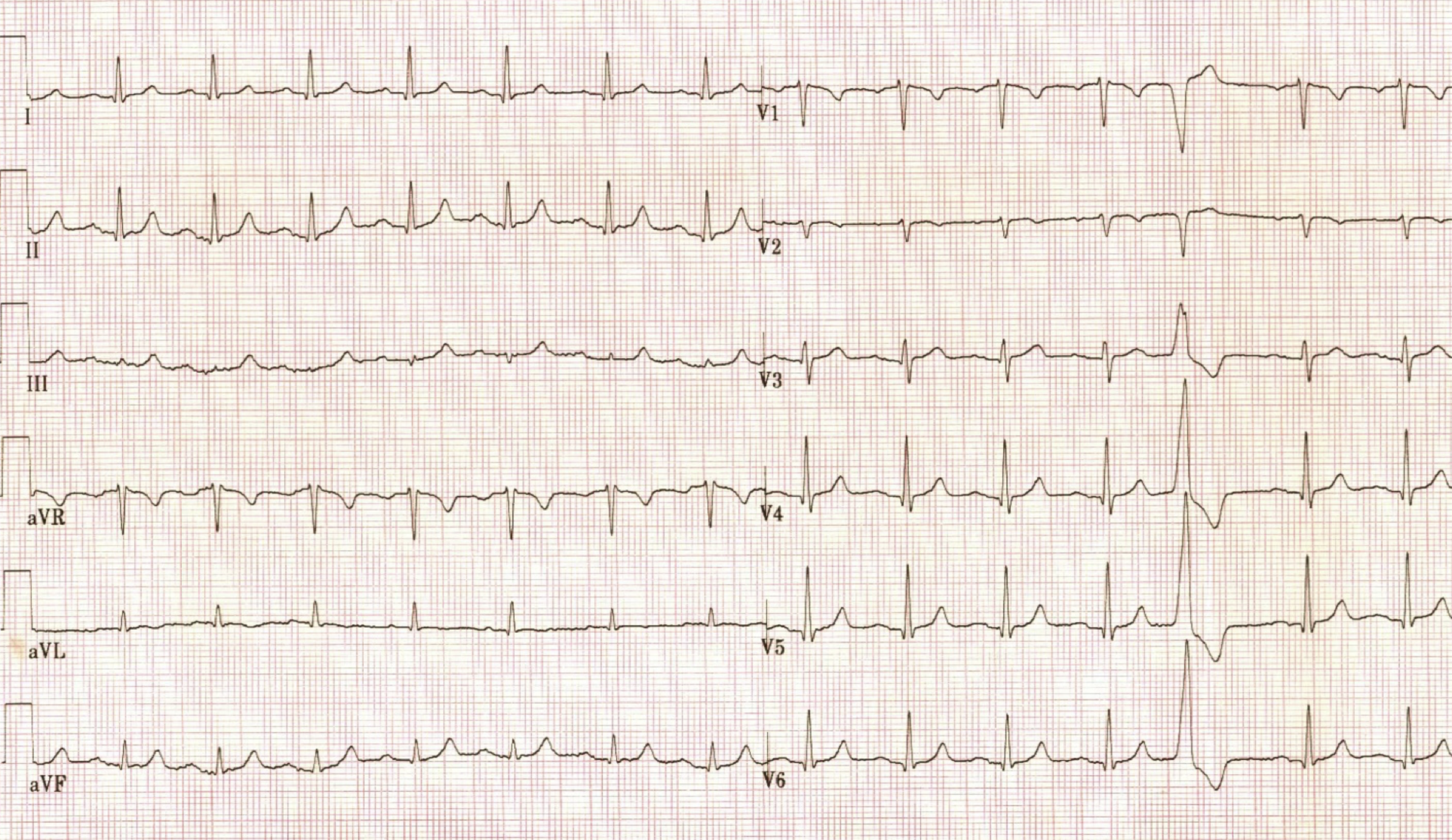}
    \caption{ECG with normofrequent sinus rhythm (heart rate 90/min), normal QRS duration and normal R wave progression. One ventricular extra beat with missing P wave and broad QRS complex with subsequent ST segment depression (V3-V6).}
    \label{fig:ecg1}
\end{figure*}

\begin{figure}
\begin{lstlisting}[style=ChronicleDB]
SELECT r_r_start, r_r_end, r_r_distance
FROM (
    SELECT r_time_1 AS r_r_start, r_time_2 AS r_r_end, 
           r_time_2 - r_time_1 AS r_r_distance
    FROM (
        SELECT r_time
        FROM myecg1
        MATCH_RECOGNIZE(
            MEASURES
                B.lead_I AS q_value,
                E.timestep AS r_time,
                E.lead_I AS r_value
            PATTERN A B C+ D+ E
            DEFINE
                A AS TRUE,
                B AS PREV(lead_I) >= lead_I,
                C AS PREV(lead_I) < lead_I,
                D AS (q_value+300) < lead_I AND PREV(lead_I) < lead_I,
                E AS (q_value+300) < lead_I AND PREV(lead_I) > lead_I
            WITHIN 50 MILLISECONDS )
        ) AS Peaks
    MATCH_RECOGNIZE(
        MEASURES
            A.r_time AS r_time_1,
            B.r_time AS r_time_2
        PATTERN
        A B DEFINE A AS TRUE, B AS TRUE )
   ) AS Distances
WINDOW (COUNT 5000 EVENTS JUMP 5000 EVENT)
\end{lstlisting}
\caption{R-R-interval query. The pattern is defined similar to a regular expression. Parts A-E are defined based on comparison of previous and current value appearing in the time series. In this example, the R-R-intervals in lead I of myecg1 are queried, resulting in a set of start- and end-points as well as distances.}
\label{fig:rrquery}
\end{figure}

\begin{figure*}
    \centering
    \includegraphics[width=\textwidth]{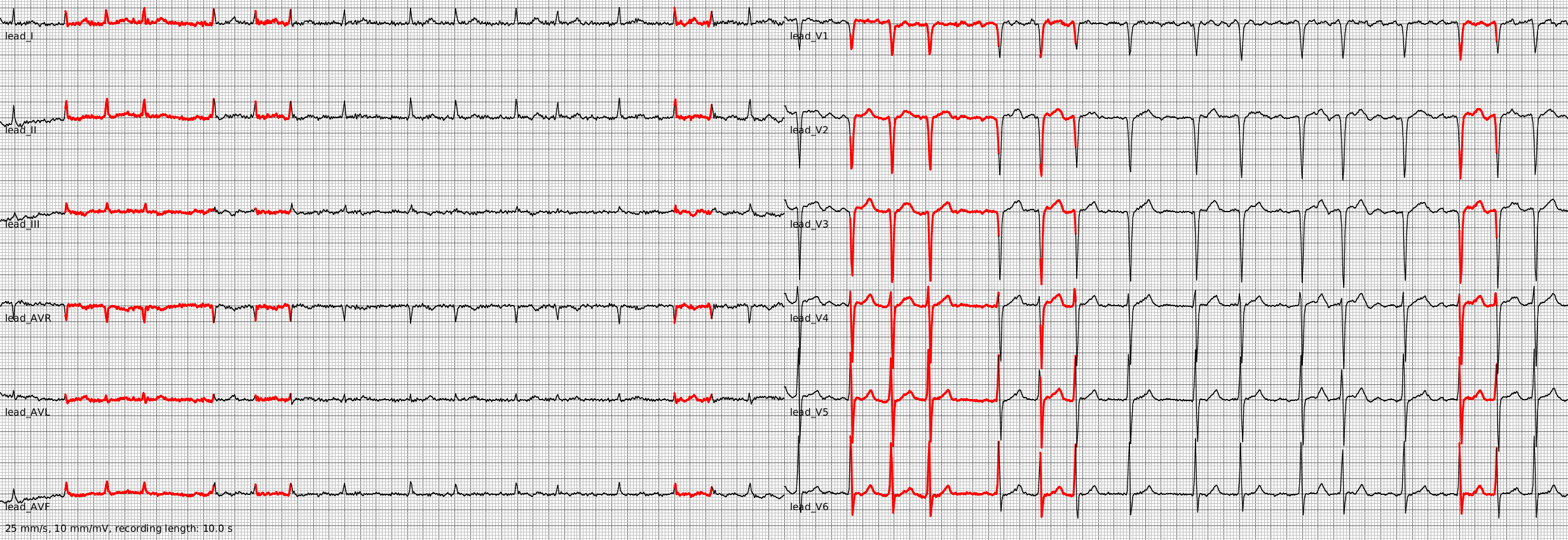}
    \caption{ECG with highlighted R-R-intervals. R-R-intervals that differ relevantly from the mean R-R-interval distance are marked bold/red. R-R-intervals were calculated based on lead I and extrapolated to the remaining leads.}
    \label{fig:ecgrrcolor}
\end{figure*}

Cardiac arrhythmia and especially atrial fibrillation (AF) are relevant health problems worldwide. This significance of AF will further grow with a current estimated prevalence of 3.0\% in Europe \citep{Kirchhof2016} and the expectation of a further rise in the upcoming years \citep{Wilke2012, Lane2017}. Individuals suffering AF have an increased risk for stroke or embolic events as well as development of heart failure \citep{Bjoerck2013, Stewart2002}. Prevention of these grave consequences of AF, based on an early identification and adequate therapy, is of great importance in health care. 

Robust, adequate and timely diagnosis of AF is often challenging as many patients with AF are asymptomatic. This lack of specific symptoms in AF is seen especially early after disease onset, hence many patients are not searching the expertise of a medical doctor specialised in cardiac diseases at an early time point. This supports the need of widely available screening tools with a low threshold for an application. The gold-standard diagnostic instrument for identification of AF is the electrocardiogram (ECG). Using an ECG to diagnose cardiac abnormalities, doctors actually try to identify internalised patterns they learned in medical school and during training. The ECG provides the human-interpretable visualisation of the underlying time series of changes in the electrical activity of the heart measured via skin electrodes. Exemplary patterns visualised by an ECG are shown in Figure 4.  

The potential applications of artificial intelligence (AI) for doctors who work with patterns and for doctors who work without patterns have recently been discussed by \citet{Topol2019}. Examples of such areas and patterns used by doctors e.g. for diagnosing can be radiologists analysing output of magnetic resonance imaging devices but also cardiologists looking for pathological patterns in ECG time series.

A system that can store and visualise ECG time series while highlighting suspicious sequences that resemble known patterns together with a human reader delivers a blueprint for a system in which healthcare professionals in general can be assisted by an algorithmic diagnosis tool providing expert -level knowledge. 

Within this second case study, we used an SQL-based query language in a time-series database (ChronicleDB \citep{Seidemann2019}) to prototypically describe a specific and common ECG pattern, the so-called R peak. This enabled us to query (see figure \ref{fig:rrquery}) all R peaks from an ECG and to calculate the respective R-R-interval distances. In healthy subjects (here with constant sinus rhythm), this R-R-interval shows only a small variance, hence the R-R-distances in an ECG are situated tightly around the mean R-R-distance. Certain diseases, concerning the heart rhythm and especially arrhythmia (e.g. based on AF) yield a large R-R-interval variance, causing a broader distribution of R-R-distances around the mean. To explore this concept, we identified such potentially pathological ECGs that showed R-R-intervals that differed relevantly from the mean, here with more than the standard deviation as threshold. 

To provide an interface for the human interpreter, these abnormal  R-R-intervals of interest were highlighted in an ECG visualisation (see figure \ref{fig:ecgrrcolor}). The visual inspection of the ECG by healthcare professionals can be  assisted by highlighting e.g. abnormal R-R-intervals. Trust is provided by showing a familiar representation i.e. all information the expert are used to have for diagnosis and, in addition, visually indicate abnormalities of common ECG patterns.

Besides rhythmic disorders, cardiac diseases that are associated with structural changes of the heart -- such as myocardial scar -- leave only scattered interference in the ECG time series, producing patterns that are, if visible at all, only accessible for well-trained expert cardiologists \citep{Markendorf2019}. However, they can be identified by an appropriately structured and sufficiently trained artificial neural network (ANN) \citep{Gumpfer2019, Gumpfer2020}. Beyond the discriminatory information itself that can be used in AI-aided diagnostic concepts, such systems further can be used to visualise areas in its input data favouring the classification made by the algorithm. Such visualisations with e.g. heat maps are a common means of explainable AI with focus on ANNs. Once explained and shown, doctors can learn and internalise those newly identified patterns.   

\begin{figure}
    \centering
    \includegraphics[width=\linewidth]{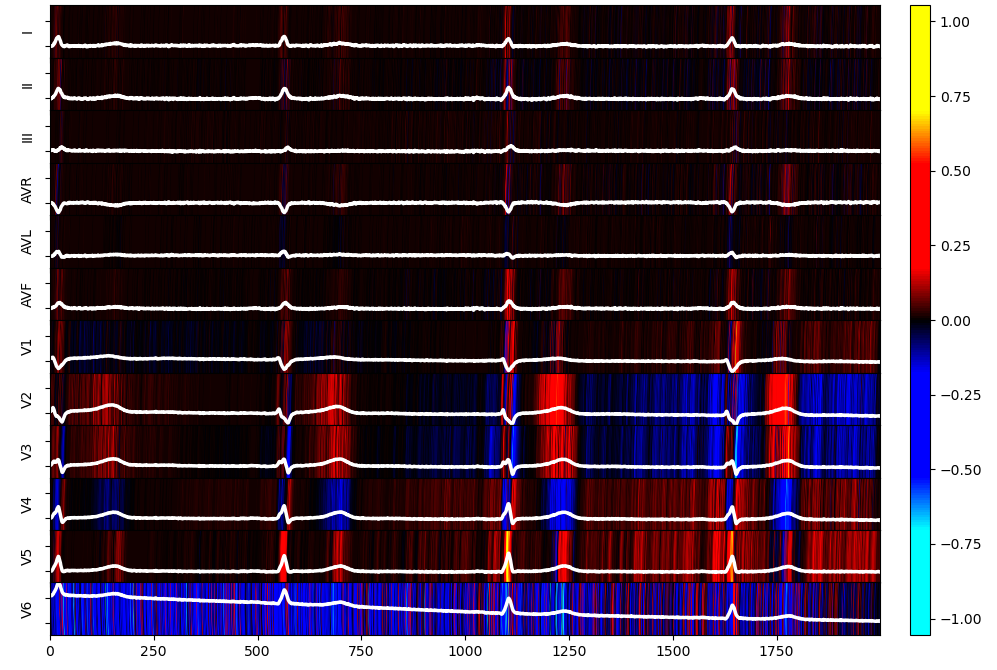}
    \caption{Attributions of ECG input on neural network output. Yellow and red parts correspond to parts that support the given prediction and blue parts correspond to parts that support disagreement with the prediction. This case is a true positive prediction. The disease of interest is a structural change in myocardium (e.g. scar).}
    \label{fig:ecgnnattributions}
\end{figure}

To highlight these parts in the ECG that had the highest attribution on the prediction of the ANN (see figure \ref{fig:ecgnnattributions}), we used Layerwise-Relevance-Propagation proposed by \citet{Bach2015}. In this exemplary heatmap, you can see that there is much focus on the R peaks (e.g. leads AVR and V5), but also on the T waves (especially lead V2). This is reasonable, as differences in height of the R peak and the T wave are known to be indicators for structural changes in the myocardium \citep[pp.11-13]{Petty2020}. Further, there is negative attribution on the prediction at the beginning of lead V6, what shows that the network indeed also relies on wrong patterns. In this very case, this pattern was caused by breathing, what changed the distance between myocardium and the electrode during recording, resulting in varying signal intensity. In AI, uncovering such behaviour is termed as \emph{unmasking Clever Hans Predictors} \citep{Lapuschkin2019}.

Such visualisations as shown in figures {\ref{fig:ecgrrcolor}} and {\ref{fig:ecgnnattributions}} can elicit trust among the healthcare professionals because all information used in the conventional, human-based diagnostic process are still visible. Based on this familiar representation, the AI can explain its prediction by contextual integration. Furthermore, confidence in AI is supported by the identification of known ECG patterns that correspond with the personal catalogue of patterns for this type of cardiac diseases.  

By feeding patterns newly identified by the AI back into an externalised catalogue of patterns to be assessed by a human expert, a learning loop can be established in which the human expert extends her/his domain knowledge by using the AI application as an assisting tool and thus perceiving it not as a potential substitute for his role but rather as an extension. 

An externalised catalogue of patterns can also be used to compose and refine training data for the AI. Simple ECG patterns can be pre-extracted by methods as described above and used for labelling or as inputs for the ANN directly. With this approach, more sophisticated AI applications can be developed, whose results may be fed back into the pattern catalogue.

%% file: patternframework.tex
The analysis of our two cases identified knowledge domains in which experts possess conceptual patterns, which they use to solve the demanding tasks of their art. We could observe this for tasks that have a high diagnostic nature, like medical diagnosis or the evaluation of complex transportation schemes. These patterns allow experts to assess assumptions and solutions for problems as appropriate and fit for the purpose. As such they manifest expectations and conceptions of the expert in his domain. 

Some tasks strongly rely on determining patterns and deriving results by evaluating and assessing these identified structures. Doctors that use data and images for their diagnosis as radiologists or cardiologists are typical examples \citep{Topol2019}. Patterns in the ECG that indicate problems with the heart are known probably since the moment the ECG recording was invented and its potential realised. However, new patterns correlating to diseases have continuously been detected by researching cardiologists and then been accepted by the community of general practitioners. This is how knowledge is extended and evolves over time.

From a knowledge management perspective, we can view patterns as manifestations of knowledge assets. As two sides of a medal we have intrinsic or internalised knowledge on one side and what is called externalised knowledge on the other. This has extensively been discussed in the past \citep{Nonaka1995}. In each of our cases, the externalisation process maps a knowledge asset onto a formalised linguistic formulation of a pattern, i.e. a regular expression or a SQL query. The formal expressions can be visualised and thus again be used for explanation, e.g. for a lay person or a student new to the field who then internalises that particular piece of knowledge.

If algorithms produce new, disruptive solutions they typically only contain few familiar patterns and potentially lots of new ones. A frequently cited example for this is Deep Mind's Alpha Go Zero \citep{Silver2017} that calculated results that even professional Go players could hardly retrace \citep{wired2018}. We therefore need instruments that give insight into innovative solutions and with which a new pattern can first be identified and then be formalised \citep{DBLP:conf/gecco/UrquhartGP19}. 

To illustrate the above we look at the following example. Assume that an experienced delivery planner on basis of  past experience believes that whenever a delivery van passes a customer,  goods should always be delivered immediately.  Let us further assume that an evolutionary algorithm has constructed a very good solution within which this is not the case and that this solution is better from the perspective of the objectives of the underlying optimisation problem. The expert will very likely reject the new solution because of the - at least for him - opaque working of the algorithm and because it contradicts his view of what is good, i.e. his patterns of a good transportation plan. 

Similarly with diagnoses produced by AI algorithms that detect patterns in an ECG time series that are not yet common knowledge of cardiologists. Again visualisation may act as a bridge and help to define the corresponding query that retrieves patterns in the time series database.  

Popper introduced the idea of three ontological worlds in his famous epistemological model \citep{Popper1972}. In this model, the first world (W 1) is the world of physical objects and real world processes. Here biology and medicine are situated. The second world (W 2) is the world of mental events, processes, and predispositions. Here we locate beliefs and psychological phenomena. The third world (W 3) is the world of the products of the human mind materialised in linguistic formulations either oral or written. \citet{McElroy2003} model the knowledge acquisition process of organisations through what they call \textit{knowledge claims}. Knowledge claims are conjectures that give tentative solutions for newly stated  problems. In accordance with Popper's tetradic schema, they can later either be refuted if they show inconsistencies (with W 1) - i.e. they are falsified - or get accepted as newly acquired knowledge if they withstand testing. This model is based on Popper's three world model and its knowledge acquisition process through trial and error. 

We will project this three world perspective onto our discussion of patterns (see figure \ref{fig:FigConcFrame}):   

\begin{figure}
    \centering
    \includegraphics[width=0.5\textwidth]{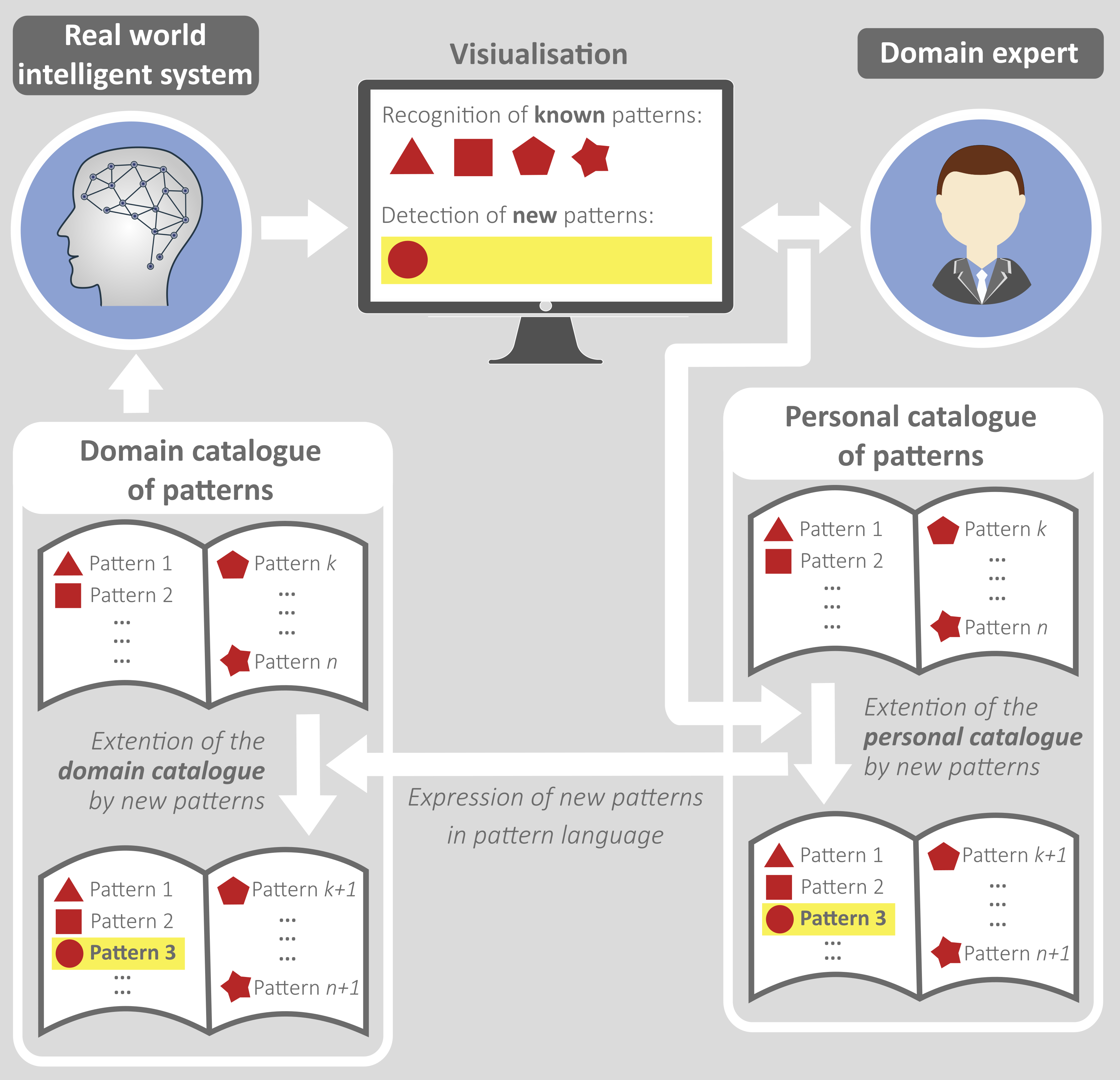}
    \caption{The conceptual framework of pattern detection to establish trust in real world intelligent systems.}
    \label{fig:FigConcFrame}
\end{figure}

Let us call the available internalised patterns the \textit{personal catalogue of patterns} of an expert and the set of externalised patterns of a domain the \textit{domain catalogue of patterns}. We see the personal catalogue as an interior manifestation of knowledge in the mind of an expert. That domain catalogue of patterns consists of knowledge patterns that can be expressed in the \textit{pattern language} of that catalogue. New patterns are analogous to the knowledge claims mentioned before. While working on the diagnosis of a complicated case or while scrutinising the solution of a complex transportation problem produced by a non-deterministic algorithm an expert will create conjectures and hypotheses that help her understand and assess the quality of the underlying matter. This will lead to observations that potentially can explain or indicate solutions. The structure of this observation can become a claim for a pattern. Analysis and application in further cases may transform the at first only conjectured pattern claim into a newly acquired knowledge pattern that is added to the individual catalogue of that expert. Eventually, it can and should be externalised and expressed in the pattern language by integrating it into the domain knowledge catalogue and thus extend it. Analogous to Popper's W 2, our conceptual framework contains the more subjective individual \textit{personal catalogue of patterns} and the externalised \textit{domain catalogue of patterns} that corresponds to W 3 which is then available to a public discourse and further testing.

Patterns are a useful means to assist in the understanding of results and diagnoses and may be a catalyst to providing explanations. Explanations are a key for fostering trust in AI systems \citep{Samek2019}. The right to explanation has even found its way into the EU's General Data Protection Regulation \citep{EUGDPR2016}. Article 13, 14 and 22  address the use of personal data and article 13 explicitly states that "meaningful information about the logic involved" shall be provided if collected data is subject to automated decision making. Moreover, by order of the German Federal Ministry of the Interior (BMI) a panel of experts has created a report seeking answers to principal ethical and legal questions of how data can and should be used. In this report explainablity, retraceability and transparency have been identified as crucial requirements for the use of data processing algorithms \citep{BMI2019}. 

We argue that there is a small but significant difference between explainability and interpretability. \citet{Doshi-Velez2017} define \textit{interpretability} in the context of machine learning as the ability to explain or to present results to a human in terms understandable for the human. This implies that interpretability uses a type of language in its presentation that is accessible for the human being. Gilpin and colleagues define interpretability and completeness as two goals of explainable AI for which an appropriate trade-off must be found \citep{DBLP:journals/corr/abs-1806-00069}. Explainability does not necessarily need to fully bridge the language gap to the user and may stay within the representation of the algorithm, e.g. be based on a visualisation of the activation layers of the neural network. According to \cite{Lipton2018} explanations will reflect the inner workings of a system or at least deliver useful information about it. Explainability is therefore a mechanism with which an intelligent systems describes how a certain result was produced e.g. by highlighting internal structures relevant for the decision. \cite{NIST2020} define five dimensions of explainability with different levels of complexity addressing different requirements and user groups. Trust is named as an outcome of understanding which again is supported by explainability.  \cite{10.1093/jamia/ocz229} conducted an exemplary survey on a group of physicians applying a ML risk calculator to solve a diagnostic dilemma. The survey showed that understanding, explainability and trust are significantly related. 
The authors argue that qualitative studies and opinion articles outline the desire of physicians to understand the logic of an intelligent algorithm before they are ready to follow its recommendations. In the case of doctors, understanding results of an intelligent algorithm depends on interpreting the results in order to produce a rationale for e.g. a diagnosis. This can obviously be supported by explanations given by the system, maybe with model agnostic explanations being preferred (again see \cite{10.1093/jamia/ocz229}). Repetitive, successful and plausible use of such an instrument will eventually create inductive trust built on this personal past experience although it could be objected that plausibility is at least partially subjective (see \cite{8558724}). Inductive trust will ideally reduce the requests for explanations to the more rare and ambiguous cases. Even, as \cite{London2019} states, doctors have by all means always been willing to accept opacity in their decision making such trust building mechanisms will help establish acceptance of intelligent systems.

We see the basic question of interpretability as \textit{Why?} and that of explainability as \textit{How?}. Our case studies show in which way the answer to \textit{Why?} can be given in a pattern language that is not within the machine learning model and can therefore be considered as being model agnostic. The answer to \textit{How?} links this interpretation to relevant structures in the ML model. However, combined answers to both questions, understandable for the user can be a key to acceptance of unusual, innovative solutions as the solutions are. Referring back to our transportation expert, explanations of how the solution was derived together with why it is a good solution, given in a language that she is able to understand will create trust in the quality of the solution and potentially lead to an eventual adaptation of the catalogues of patterns. 

Considering supervised learning which is the case in our second case study we can speak of a form of induction, i.e. of drawing conclusions about the future based on past experience. Popper anticipated long before us thinking about intelligent systems an "induction machine" that applied to sequences of coloured counters "...may through repetition 'learn', or even 'formulate', laws of succession" while neglecting pure induction as source of scientific knowledge. The framework presented here helps in creating hypotheses based on observations (e.g. patterns in the ECG) of W1 objects. These hypotheses formulated as patterns can undergo inspection and test before being accepted. Trust in the machines can be established and deepened by such a learning process in which users can gradually adapt their expectations to the results explained by the algorithm. By using intelligent systems they get new insights into their domain and build new conceptions of it. Algorithmic results do not have to be simply accepted but can be scrutinised until they are sufficiently illuminated.

The choice of our case studies shows that we focus on semi-autonomous systems in which the intelligent algorithms supports the user in his decision. We do not aim at fully automated systems. However, patterns can potentially be used in settings with higher degrees of independence. For example, tested queries can be used to monitor ECG recordings collected by wearable devices. Here, again, we have an analogy to Popper's knowledge acquisition model: a testified query has become an accepted and applicable knowledge asset, which can be used until further refined or replaced.

%% file: relatedwork.tex
Our conceptual framework can potentially be applied to a wide variety of algorithms, e.g. deep learning or evolutionary algorithms as demonstrated by our case studies.
In this context, methods of explainability are a valuable tool with which relevant patterns can be identified and be made explicit.

In the area of AI-based ECG processing, different explainability approaches have been proposed. \citet{Teijeiro2018} used a knowledge-based algorithm relying on abductive reasoning to train their neural network, what enables explainability by design. \citet{Strodthoff2019} have shown that recent explainable AI methods \citep{Shrikumar2017, Bach2015, Sundararajan2017} are applicable to different kinds of neural networks trained on ECGs. 

Illumination algorithms create series of possible solutions. From the perspective of the expert for evolutionary algorithms they illuminate how the algorithm got to the solution and from the perspective of the domain expert they present results along the dimensions defined by the objectives of the optimisation problem \citep{Mouret-2015}. 

The field of Interactive Machine Learning (IML) has the goal to define machine learning algorithms that incorporate meaningful interaction with humans. Dudley and Kristenson see IML as a paradigm in which input and review by humans iteratively refine the mathematical model constructed by the learning algorithm \citep{Dudley2018}. Moreover, Active Learning requires an oracle, i.e. a human interacting with the system, to label relevant data points that are not appropriately labelled to be used in the learning algorithm. Our paradigm of the catalogue of patterns also addresses human experts to interact with our system. But different from IML approaches we do not expect the user to intervene in the processes of the algorithm itself. We aim at assisting the user in managing his routine tasks with her routine assumptions together with the opportunity to create new ways of problem solving and diagnosis.   

Abdel-Karim and colleagues focus on human learning while being in the loop of human-machine interaction in interactive machine learning \citep{Abdel-Karim2020}. They propose a contradiction matrix which similarly to a traditional confusion matrix documents the number of cases in which human and machine diagnosis are equivalent and in which they are equal. Similar to our ideas this instrument can be considered as a knowledge management instrument. Interactive explanations with WhatIf-questions have been examined to create glass-box AI systems \citep{Sokol2020}.

However, we see our pattern based approach less as an instrument of IML but more as a move towards explainability and interpretability. Using queries and patterns in a more dialogue based form are a potential strand of future extensions.   

%% file: conclusion.tex
We have presented an argument that trust in intelligent systems may potentially be engendered through the use of tools that formalise and visualise patterns that are integral components of the way experts in the specific field think and argue. 
Two case studies have been discussed in which users work with intelligent systems. One being the MAP-Elites algorithm that can help to explain the operation of stochastic population-based optimisation algorithms to users. The other, an ANN diagnosing heart diseases in ECG times series producing visualisations of its internal inference process.

Future work directions are a formalisation of the catalogue of patterns and how it evolves. A second research topic lies in the analysis of how users interact with tools like those presented in the case studies of this paper. This interactions may give new insights in how new knowledge is acquired and how the formal representation of patterns and the corresponding search process can be automated. Empirical studies with experts in the fields will extend the respective catalogues and will deliver data to formalise the interactions as well as show the scalability of the approach.    

%% file: main.bbl
\begin{thebibliography}{45}
\providecommand{\natexlab}[1]{#1}
\providecommand{\url}[1]{\texttt{#1}}
\providecommand{\urlprefix}{URL }
\expandafter\ifx\csname urlstyle\endcsname\relax
  \providecommand{\doi}[1]{doi:\discretionary{}{}{}#1}\else
  \providecommand{\doi}{doi:\discretionary{}{}{}\begingroup
  \urlstyle{rm}\Url}\fi

\bibitem[{Abdel-Karim et~al.(2020)Abdel-Karim, Pfeuffer, and
  Rohde}]{Abdel-Karim2020}
Abdel-Karim, B., Pfeuffer, N., and Rohde, G. (2020).
\newblock How and what can humans learn from being in the loop?
\newblock \emph{K{\"u}nstl Intell}, 34, 199--207.
\newblock \doi{10.1007/s13218-020-00638-x}.

\bibitem[{{Andras} et~al.(2018){Andras}, {Esterle}, {Guckert}, {Han}, {Lewis},
  {Milanovic}, {Payne}, {Perret}, {Pitt}, {Powers}, {Urquhart}, and
  {Wells}}]{8558724}
{Andras}, P., {Esterle}, L., {Guckert}, M., {Han}, T.A., {Lewis}, P.R.,
  {Milanovic}, K., {Payne}, T., {Perret}, C., {Pitt}, J., {Powers}, S.T.,
  {Urquhart}, N., and {Wells}, S. (2018).
\newblock Trusting intelligent machines: Deepening trust within socio-technical
  systems.
\newblock \emph{IEEE Technology and Society Magazine}, 37(4), 76--83.

\bibitem[{Bach et~al.(2015)Bach, Binder, Montavon, Klauschen, Müller, and
  Samek}]{Bach2015}
Bach, S., Binder, A., Montavon, G., Klauschen, F., Müller, K.R., and Samek, W.
  (2015).
\newblock On pixel-wise explanations for non-linear classifier decisions by
  layer-wise relevance propagation.
\newblock \emph{PLOS ONE}, 10(7), 1--46.
\newblock \doi{10.1371/journal.pone.0130140}.

\bibitem[{Bjoerck et~al.(2013)Bjoerck, Palaszewski, Friberg, and
  Bergfeldt}]{Bjoerck2013}
Bjoerck, S., Palaszewski, B., Friberg, L., and Bergfeldt, L. (2013).
\newblock Atrial fibrillation, stroke risk, and warfarin therapy revisited.
\newblock \emph{Stroke}, 44(11), 3103--3108.
\newblock \doi{10.1161/STROKEAHA.113.002329}.

\bibitem[{Cummings(2004)}]{Cummings04automationbias}
Cummings, M.L. (2004).
\newblock Automation bias in intelligent time critical decision support
  systems.
\newblock In \emph{AIAA 3rd Intelligent Systems Conference}, 2004--6313. AIAA.

\bibitem[{Datenethikkommission(2019)}]{BMI2019}
Datenethikkommission (2019).
\newblock Gutachten der datenethikkommission.
\newblock
  \urlprefix\url{https://www.bmi.bund.de/SharedDocs/downloads/DE/publikationen/themen/it-digitalpolitik/gutachten-datenethikkommission.pdf}.

\bibitem[{Diprose et~al.(2020)Diprose, Buist, Hua, Thurier, Shand, and
  Robinson}]{10.1093/jamia/ocz229}
Diprose, W.K., Buist, N., Hua, N., Thurier, Q., Shand, G., and Robinson, R.
  (2020).
\newblock {Physician understanding, explainability, and trust in a hypothetical
  machine learning risk calculator}.
\newblock \emph{Journal of the American Medical Informatics Association},
  27(4), 592--600.
\newblock \doi{10.1093/jamia/ocz229}.
\newblock \urlprefix\url{https://doi.org/10.1093/jamia/ocz229}.

\bibitem[{Doshi-Velez and Kim(2017)}]{Doshi-Velez2017}
Doshi-Velez, F. and Kim, B. (2017).
\newblock Towards a rigorous science of interpretable machine learning.
\newblock \emph{CoRR}, abs/1702.08608.
\newblock \urlprefix\url{http://arxiv.org/abs/1702.08608}.

\bibitem[{Dudley(2018)}]{Dudley2018}
Dudley, J.;~Kristenson, P. (2018).
\newblock A review of user interface design for interactive machine learning.
\newblock \emph{The ACM Transactions on Interactive Intelligent Systems}, 8(2),
  1--37.
\newblock \doi{10.1145/3185517}.

\bibitem[{{European Parliament and the Council of the European
  Union}(2016)}]{EUGDPR2016}
{European Parliament and the Council of the European Union} (2016).
\newblock {REGULATION (EU) 2016/679 OF THE EUROPEAN PARLIAMENT AND OF THE
  COUNCIL of 27 April 2016 on the protection of natural persons with regard to
  the processing of personal data and on the free movement of such data, and
  repealing Directive 95/46/EC (General Data Protection Regulation)}.
\newblock \emph{Official Journal of the European Union}, L119/1.
\newblock
  \urlprefix\url{https://eur-lex.europa.eu/legal-content/EN/TXT/PDF/?uri=CELEX:32016R0679}.

\bibitem[{Firestone and McElroy(2003)}]{McElroy2003}
Firestone, J.M. and McElroy, M.W. (2003).
\newblock \emph{Key Issues in the New Knowledge Management}.
\newblock KMCI Press. Butterworth-Heinemann, Boston.
\newblock \doi{10.1016/B978-0-7506-7655-7.50012-2}.

\bibitem[{Gilpin et~al.(2018)Gilpin, Bau, Yuan, Bajwa, Specter, and
  Kagal}]{DBLP:journals/corr/abs-1806-00069}
Gilpin, L.H., Bau, D., Yuan, B.Z., Bajwa, A., Specter, M., and Kagal, L.
  (2018).
\newblock Explaining explanations: An approach to evaluating interpretability
  of machine learning.
\newblock \emph{CoRR}, abs/1806.00069.
\newblock \urlprefix\url{http://arxiv.org/abs/1806.00069}.

\bibitem[{Gumpfer et~al.(2020)Gumpfer, Grün, Hannig, Keller, and
  Guckert}]{Gumpfer2020}
Gumpfer, N., Grün, D., Hannig, J., Keller, T., and Guckert, M. (2020).
\newblock Detecting myocardial scar using electrocardiogram data and deep
  neural networks.
\newblock \emph{Biological chemistry}.
\newblock \doi{10.1515/hsz-2020-0169}.

\bibitem[{Gumpfer et~al.(2019)Gumpfer, Hurka, Hannig, Rolf, Hamm, Guckert, and
  Keller}]{Gumpfer2019}
Gumpfer, N., Hurka, S., Hannig, J., Rolf, A., Hamm, C.W., Guckert, M., and
  Keller, T. (2019).
\newblock Development of a machine learning algorithm to predict myocardial
  scar based on a 12-lead electrocardiogram.
\newblock \emph{Clinical Research in Cardiology}, 108(1).
\newblock \doi{10.1007/s00392-019-1435-9}.
\newblock V1802.

\bibitem[{Hart and Ross(2001)}]{hart_ross_2001}
Hart, E. and Ross, P. (2001).
\newblock Gavel-a new tool for genetic algorithm visualization.
\newblock \emph{IEEE Transactions on Evolutionary Computation}, 5(4), 335--348.

\bibitem[{Holland et~al.(1992)}]{holland_1992}
Holland, J.H. et~al. (1992).
\newblock \emph{Adaptation in natural and artificial systems: an introductory
  analysis with applications to biology, control, and artificial intelligence}.
\newblock MIT press.

\bibitem[{Inselberg(2009)}]{inselberg_shneiderman_2009}
Inselberg, A. (2009).
\newblock \emph{Parallel coordinates: visual multidimensional geometry and its
  applications}, volume~20.
\newblock Springer Science \& Business Media.
\newblock \doi{10.1007/978-0-387-68628-8}.

\bibitem[{Kirchhof et~al.(2016)Kirchhof, Benussi, Kotecha, Ahlsson, Atar,
  Casadei, Castella, Diener, Heidbuchel, Hendriks, Hindricks, Manolis, Oldgren,
  Popescu, Schotten, Van~Putte, Vardas, and Group}]{Kirchhof2016}
Kirchhof, P., Benussi, S., Kotecha, D., Ahlsson, A., Atar, D., Casadei, B.,
  Castella, M., Diener, H.C., Heidbuchel, H., Hendriks, J., Hindricks, G.,
  Manolis, A.S., Oldgren, J., Popescu, B.A., Schotten, U., Van~Putte, B.,
  Vardas, P., and Group, E.S.D. (2016).
\newblock {2016 ESC Guidelines for the management of atrial fibrillation
  developed in collaboration with EACTS}.
\newblock \emph{European Heart Journal}, 37(38), 2893--2962.
\newblock \doi{10.1093/eurheartj/ehw210}.

\bibitem[{Lane et~al.(2017)Lane, Skjoth, Lip, Larsen, and Kotecha}]{Lane2017}
Lane, D.A., Skjoth, F., Lip, G.Y.H., Larsen, T.B., and Kotecha, D. (2017).
\newblock Temporal trends in incidence, prevalence, and mortality of atrial
  fibrillation in primary care.
\newblock \emph{Journal of the American Heart Association}, 6(5), e005155.
\newblock \doi{10.1161/JAHA.116.005155}.

\bibitem[{Lapuschkin et~al.(2019)Lapuschkin, W{\"a}ldchen, Binder, Montavon,
  Samek, and M{\"u}ller}]{Lapuschkin2019}
Lapuschkin, S., W{\"a}ldchen, S., Binder, A., Montavon, G., Samek, W., and
  M{\"u}ller, K.R. (2019).
\newblock Unmasking clever hans predictors and assessing what machines really
  learn.
\newblock \emph{Nature Communications}, 10(1), 1096.
\newblock \doi{10.1038/s41467-019-08987-4}.

\bibitem[{Liao(2002)}]{Liao2002}
Liao, S. (2002).
\newblock Problem solving and knowledge inertia.
\newblock \emph{Expert Systems with Applications}, 1, 21--31.
\newblock \doi{10.1016/S0957-4174(01)00046-X}.

\bibitem[{Lipton(2018)}]{Lipton2018}
Lipton, Z.C. (2018).
\newblock The mythos of model interpretability: In machine learning, the
  concept of interpretability is both important and slippery.
\newblock \emph{Queue}, 16(3), 31–57.
\newblock \doi{10.1145/3236386.3241340}.
\newblock \urlprefix\url{https://doi.org/10.1145/3236386.3241340}.

\bibitem[{London(2019)}]{London2019}
London, A. (2019).
\newblock Artificial intelligence and black-box medical decisions: Accuracy
  versus explainability.
\newblock \emph{Hastings Cent Rep}, 49(1), 15--21.
\newblock \doi{doi: 10.1002/hast.973.}

\bibitem[{Markendorf et~al.(2019)Markendorf, Benz, Messerli, Grossmann,
  Giannopoulos, Patriki, Fuchs, Gr{\"a}ni, Pazhenkottil, Buechel, Kaufmann, and
  Gaemperli}]{Markendorf2019}
Markendorf, S., Benz, D.C., Messerli, M., Grossmann, M., Giannopoulos, A.A.,
  Patriki, D., Fuchs, T.A., Gr{\"a}ni, C., Pazhenkottil, A.P., Buechel, R.R.,
  Kaufmann, P.A., and Gaemperli, O. (2019).
\newblock Value of 12-lead electrocardiogram to predict myocardial scar on fdg
  pet in heart failure patients.
\newblock \emph{Journal of Nuclear Cardiology}.
\newblock \doi{10.1007/s12350-019-01841-6}.

\bibitem[{Mouret and Clune(2015)}]{Mouret-2015}
Mouret, J. and Clune, J. (2015).
\newblock Illuminating search spaces by mapping elites.
\newblock \emph{CoRR}, abs/1504.04909.
\newblock \urlprefix\url{http://arxiv.org/abs/1504.04909}.

\bibitem[{Nonaka and Takeuchi(1995)}]{Nonaka1995}
Nonaka, I. and Takeuchi, H. (1995).
\newblock \emph{The Knowledge-Creating Company: How Japanese Companies Create
  the Dynamics of Innovation}.
\newblock Oxford University Press, New York.

\bibitem[{Pallier et~al.(2002)Pallier, Wilkinson, and Roberts}]{Pallier2002}
Pallier, G., Wilkinson, R., and Roberts, R. (2002).
\newblock The role of individual differences in the accuracy of confidence
  judgments.
\newblock \emph{The Journal of General Psychology}.
\newblock \doi{10.1080/00221300209602099}.

\bibitem[{Petty(2020)}]{Petty2020}
Petty, B.G. (2020).
\newblock \emph{Basic Electrocardiography}.
\newblock Springer, New York, NY.
\newblock \doi{10.1007/978-1-4939-2413-4}.

\bibitem[{Phillips et~al.(2020)Phillips, Hahn, Fontana, Broniatowski, and
  Przybocki}]{NIST2020}
Phillips, P.J., Hahn, C.A., Fontana, P.C., Broniatowski, D.A., and Przybocki,
  M.A. (2020).
\newblock Four principles of explainable artificial intelligence.
\newblock \doi{https://doi.org/10.6028/NIST.IR.8312-draft}.

\bibitem[{Popper(1972)}]{Popper1972}
Popper, K.R. (1972).
\newblock \emph{Objective Knowledge}.
\newblock Oxford University Press, New York.
\newblock \doi{10.2307/2184085}.

\bibitem[{Samek and M{\"u}ller(2019)}]{Samek2019}
Samek, W. and M{\"u}ller, K.R. (2019).
\newblock Towards explainable artificial intelligence.
\newblock In \emph{Explainable AI: interpreting, explaining and visualizing
  deep learning}, 5--22. Springer.
\newblock \doi{10.1007/978-3-030-28954-6\_1}.

\bibitem[{Seidemann et~al.(2019)Seidemann, Glombiewski, K\"{o}rber, and
  Seeger}]{Seidemann2019}
Seidemann, M., Glombiewski, N., K\"{o}rber, M., and Seeger, B. (2019).
\newblock Chronicledb: A high-performance event store.
\newblock \emph{ACM Trans. Database Syst.}, 44(4).
\newblock \doi{10.1145/3342357}.
\newblock \urlprefix\url{https://doi.org/10.1145/3342357}.

\bibitem[{Shrikumar et~al.(2017)Shrikumar, Greenside, and
  Kundaje}]{Shrikumar2017}
Shrikumar, A., Greenside, P., and Kundaje, A. (2017).
\newblock Learning important features through propagating activation
  differences.
\newblock In \emph{Proceedings of the 34th International Conference on Machine
  Learning, {ICML} 2017, Sydney, NSW, Australia, 6-11 August 2017}, 3145--3153.

\bibitem[{Silver et~al.(2017)Silver, Schrittwieser, Simonyan, Antonoglou,
  Huang, Guez, Hubert, Baker, Lai, Bolton, Chen, Lillicrap, Hui, Sifre, van~den
  Driessche, Graepel, and Hassabis}]{Silver2017}
Silver, D., Schrittwieser, J., Simonyan, K., Antonoglou, I., Huang, A., Guez,
  A., Hubert, T., Baker, L., Lai, M., Bolton, A., Chen, Y., Lillicrap, T., Hui,
  F., Sifre, L., van~den Driessche, G., Graepel, T., and Hassabis, D. (2017).
\newblock Mastering the game of go without human knowledge.
\newblock \emph{Nature}, 550(7676), 354--359.
\newblock \doi{10.1038/nature24270}.

\bibitem[{Sokol and Flach(2020)}]{Sokol2020}
Sokol, K. and Flach, P. (2020).
\newblock One explanation does not fit all.
\newblock \emph{Künstl Intell}, 34, 235--250.
\newblock \doi{10.1007/s13218-020-00637-y}.

\bibitem[{Stewart et~al.(2002)Stewart, Hart, Hole, and McMurray}]{Stewart2002}
Stewart, S., Hart, C.L., Hole, D.J., and McMurray, J.J. (2002).
\newblock A population-based study of the long-term risks associated with
  atrial fibrillation: 20-year follow-up of the renfrew/paisley study.
\newblock \emph{The American Journal of Medicine}, 113(5), 359--364.
\newblock \doi{10.1016/S0002-9343(02)01236-6}.

\bibitem[{Strodthoff and Strodthoff(2019)}]{Strodthoff2019}
Strodthoff, N. and Strodthoff, C. (2019).
\newblock Detecting and interpreting myocardial infarction using fully
  convolutional neural networks.
\newblock \emph{Physiological Measurement}, 40(1), 015001.
\newblock \doi{10.1088/1361-6579/aaf34d}.

\bibitem[{Sundararajan et~al.(2017)Sundararajan, Taly, and
  Yan}]{Sundararajan2017}
Sundararajan, M., Taly, A., and Yan, Q. (2017).
\newblock Axiomatic attribution for deep networks.
\newblock In D.~Precup and Y.W. Teh (eds.), \emph{Proceedings of the 34th
  International Conference on Machine Learning}, volume~70 of \emph{Proceedings
  of Machine Learning Research}, 3319--3328. PMLR, International Convention
  Centre, Sydney, Australia.

\bibitem[{Teijeiro et~al.(2018)Teijeiro, Garc{\'{\i}}a, Castro, and
  F{\'{e}}lix}]{Teijeiro2018}
Teijeiro, T., Garc{\'{\i}}a, C.A., Castro, D., and F{\'{e}}lix, P. (2018).
\newblock Abductive reasoning as a basis to reproduce expert criteria in {ECG}
  atrial fibrillation identification.
\newblock \emph{Physiological Measurement}, 39(8), 084006.
\newblock \doi{10.1088/1361-6579/aad7e4}.

\bibitem[{Topol(2019)}]{Topol2019}
Topol, E. (2019).
\newblock \emph{Deep Medicine: How Artificial Intelligence Can Make Healthcare
  Human Again}.
\newblock Basic Books, New York.

\bibitem[{Urquhart(2019)}]{urquhart_2019el}
Urquhart, N. (2019).
\newblock Elvis (elite visualisation).
\newblock \urlprefix\url{commmute.napier.ac.uk/upload}.

\bibitem[{Urquhart et~al.(2019{\natexlab{a}})Urquhart, Guckert, and
  Powers}]{DBLP:conf/gecco/UrquhartGP19}
Urquhart, N., Guckert, M., and Powers, S.T. (2019{\natexlab{a}}).
\newblock Increasing trust in meta-heuristics by using map-elites.
\newblock In M.~L{\'{o}}pez{-}Ib{\'{a}}{\~{n}}ez, A.~Auger, and
  T.~St{\"{u}}tzle (eds.), \emph{Proceedings of the Genetic and Evolutionary
  Computation Conference Companion, {GECCO} 2019, Prague, Czech Republic, July
  13-17, 2019}, 1345--1348. {ACM}.
\newblock \doi{10.1145/3319619.3326816}.

\bibitem[{Urquhart et~al.(2019{\natexlab{b}})Urquhart, H{\"o}hl, and
  Hart}]{urquhart2019}
Urquhart, N., H{\"o}hl, S., and Hart, E. (2019{\natexlab{b}}).
\newblock An illumination algorithm approach to solving the micro-depot routing
  problem.
\newblock In \emph{Proceedings of the Genetic and Evolutionary Computation
  Conference}, 1347--1355.
\newblock \doi{10.1145/3321707.3321767}.

\bibitem[{Wilke et~al.(2012)Wilke, Groth, Mueller, Pfannkuche, Verheyen,
  Linder, Maywald, Bauersachs, and Breithardt}]{Wilke2012}
Wilke, T., Groth, A., Mueller, S., Pfannkuche, M., Verheyen, F., Linder, R.,
  Maywald, U., Bauersachs, R., and Breithardt, G. (2012).
\newblock {Incidence and prevalence of atrial fibrillation: an analysis based
  on 8.3 million patients}.
\newblock \emph{EP Europace}, 15(4), 486--493.
\newblock \doi{10.1093/europace/eus333}.

\bibitem[{WIRED(2018)}]{wired2018}
WIRED (2018).
\newblock How google’s ai viewed the move no human could understand.
\newblock
  \urlprefix\url{https://www.wired.com/2016/03/googles-ai-viewed-move-no-human-understand/}.

\end{thebibliography}
